\documentclass[12pt]{article}

\usepackage{amsmath,amssymb,amsfonts,amsthm,hyperref,bbm,latexsym,epsfig}
\usepackage{graphicx,multirow}

\newcommand{\bs}{\begin{subequations}}
\newcommand{\es}{\end{subequations}}
\newcommand{\be}{\begin{equation}}
\newcommand{\ee}{\end{equation}}
\newcommand{\ba}{\begin{eqnarray}}
\newcommand{\ea}{\end{eqnarray}}
\newcommand{\no}{\nonumber \\}

\newcommand{\zz}{\mathbbm{Z}}
\newcommand{\bone}{\mathbbm{1}}
\newcommand{\diag}{\mbox{diag}}
\newcommand{\mnu}{\mathcal{M}_\nu}

\allowdisplaybreaks

\usepackage{xcolor}

\begin{document}

\title{
\normalsize \hfill UWThPh-2016-1
\\[.5mm]
\normalsize \hfill CFTP/16-004
\\[4mm]
\LARGE Scotogenic model for co-bimaximal mixing}

\author{
P.\ M.\ Ferreira,$^{(1,2)}$\thanks{E-mail: \tt pmmferreira@fc.ul.pt}
\addtocounter{footnote}{2}
W.~Grimus,$^{(3)}$\thanks{E-mail: \tt walter.grimus@univie.ac.at}
\
D.~Jur\v{c}iukonis,$^{(4)}$\thanks{E-mail: \tt darius.jurciukonis@tfai.vu.lt}
\ and
L.~Lavoura$^{(5)}$\thanks{E-mail: \tt balio@cftp.tecnico.ulisboa.pt}
\\*[3mm]
$^{(1)} \! $
\small Instituto Superior de Engenharia de Lisboa,
\small Instituto Polit\'ecnico de Lisboa, Portugal
\\[2mm]
$^{(2)} \! $
\small CFTC,
\small Faculdade de Ci\^encias, Universidade de Lisboa, Portugal
\\[2mm]
$^{(3)} \! $
\small University of Vienna, Faculty of Physics,
\small Boltzmanngasse 5, A--1090 Wien, Austria
\\[2mm]
$^{(4)} \! $
\small University of Vilnius,
\small Institute of Theoretical Physics and Astronomy, Lithuania
\\[2mm]
$^{(5)} \! $
\small CFTP, Instituto Superior T\'ecnico, Universidade de Lisboa, Portugal
\\*[2mm]
}

\date{\today}

\maketitle

\begin{abstract}
We present a scotogenic model,
\textit{i.e.}\ a one-loop neutrino mass model
with dark right-handed neutrino gauge singlets
and one inert dark scalar gauge doublet $\eta$,
which has symmetries that lead to co-bimaximal mixing,
\textit{i.e.}\ to an atmospheric mixing angle $\theta_{23} = 45^\circ$
and to a $CP$-violating phase $\delta = \pm \pi/2$,
while the mixing angle $\theta_{13}$ remains arbitrary.
The symmetries consist of softly broken lepton numbers
$L_\alpha$ ($\alpha = e,\mu,\tau$),
a non-standard $CP$ symmetry,
and three $\zz_2$ symmetries.
We indicate two possibilities for extending the model to the quark sector.
Since the model has,
besides $\eta$,
three scalar gauge doublets,
we perform a thorough discussion of its scalar sector.
We demonstrate that it can accommodate a Standard Model-like scalar
with mass $125\, \mathrm{GeV}$,
with all the other charged and neutral scalars having much higher masses.
\end{abstract}

\newpage

\section{Introduction}

With the experimental finding that
the lepton mixing angle $\theta_{13}$ is nonzero,
many theoretical neutrino mass models fell into disfavour.
An exception is the model in ref.~\cite{GL2003},
in which there is a relation
\be
\label{sms}
S \mnu S = \mnu^*,
\quad \mbox{where}\
S = \left( \begin{array}{ccc} 1&0&0 \\ 0&0&1 \\ 0&1&0 \end{array} \right)
\ee
and $\mnu$ is the (symmetric) light-neutrino Majorana mass matrix
in the basis where the charged-lepton mass matrix is
$\diag \left( m_e, m_\mu, m_\tau \right)$.
The condition~(\ref{sms}) leads to $\sin^2{\theta_{23}} = 1/2$
and (provided $\sin{\theta_{13}} \neq 0$) $\cos\delta = 0$,
which is in agreement with the phenomenology~\cite{Forero:2014bxa};
this situation has recently been dubbed
`co-bimaximal mixing'~\cite{ma2015}.\footnote{A different way
for obtaining co-bimaximal mixing,
not involving the condition~(\ref{sms}),
has been recently proposed in ref.~\cite{rode}.}
A relevant point is that
the condition~(\ref{sms}) does not restrict the neutrino masses;
it only restricts lepton mixing.
Actually,
as a consequence of the condition~(\ref{sms}),
the lepton mixing matrix has the form~\cite{GL2003}
\be
\label{u}
U = \left( \begin{array}{ccc}
u_1 \eta_1 & u_2 \eta_2 & u_3 \eta_3 \\
w_1 \eta_1 & w_2 \eta_2 & w_3 \eta_3 \\
w_1^\ast \eta_1 & w_2^\ast \eta_2 & w_3^\ast \eta_3
\end{array} \right)
\ee
with $u_j \geq 0$,
$\left| w_j \right|^2 = \left. \left( 1 - u_j^2 \right) \right/ 2$,
and $\eta_j^2 = \pm 1$ for $j = 1, 2, 3$.
It is clear from equation~(\ref{u}) that
$\left| U_{\mu j} \right| = \left| U_{\tau j} \right|\
\forall\: j = 1, 2, 3$~\cite{fukuura}.
Note that the condition~(\ref{sms}) entails three restrictions on lepton mixing:
\begin{enumerate}
\item The atmospheric mixing angle $\theta_{23}$ is maximal,
\textit{i.e.}\ $\sin{\left( 2 \theta_{23} \right)} = 1$.
\item The $CP$-violating phase $\delta$ is $\pm \pi/2$.
\item The Majorana phase factors
in effective neutrinoless $\beta \beta$ decay are $\pm 1$.
\end{enumerate}
Because the predictions of condition~(\ref{sms})
do not depend on the neutrino masses,
it is possible that,
in some multi-Higgs-doublet models,
co-bimaximal mixing is not disturbed by the one-loop corrections
to the neutrino mass matrix~\cite{GL2002}.
This may,
in particular,
be the case in a `scotogenic' model~\cite{ma2006}.
In such a model,
the masses of the light neutrinos have radiative origin and
the particles in the loop that generates them
belong to the dark sector of astrophysics,
\textit{i.e.}\ they are prevented from mixing with the ordinary particles
by an unbroken (usually $\zz_2$) symmetry.

The purpose of this paper is to propose
a scotogenic model for the neutrino masses
which predicts co-bimaximal mixing.\footnote{Recently,
another such model,
but which employes a completely different mechanism,
has been proposed in ref.~\cite{ma2016}.
The model of ref.~\cite{ma2016}
is more complicated than the one presented in this paper
for several reasons:
(1) It has two types of dark matter,
one of them protected by a $U(1)$ symmetry
and the other one by a $\zz_2$ symmetry.
(2) It has several more fields in the dark sector.
(3) The masses of the charged leptons are of radiative origin,
just as those of the neutrinos.
(4) The soft breaking of the symmetries occurs in two steps,
with an $A_4$ symmetry in the dimension-four terms
being softly broken to $\zz_3$ through dimension-three terms
and that $\zz_3$ being softly broken through dimension-two terms.}
This is done in section~\ref{model}.
In section~\ref{quarks} we expose
two possible extensions of that model to the quark sector.
An analysis of the scalar potential of the model and of its compatibility
with the recently discovered scalar of mass $125\, \mathrm{GeV}$
is performed in section~\ref{Higgs}.
We summarize our findings in section~\ref{summary}.
Appendix~A collects some formulae from ref.~\cite{GL2002}
which are used in section~\ref{model}.

\section{The model for the lepton sector}
\label{model}

Our model is an extension of the Standard Model
with gauge symmetry $SU(2) \times U(1)$.
The usual fermion multiplets are three $D_{\alpha L}$ and three $\alpha_R$
($\alpha = e, \mu, \tau$).
Besides,
we introduce three right-handed neutrinos $\nu_{\alpha R}$;
they belong to the dark sector of the model.
Our model has four scalar doublets:
\be
\phi_j = \left( \begin{array}{c}
\phi_j^+ \\*[1mm] \phi_j^0 \end{array} \right) \ (j = 1, 2, 3)
\quad \mbox{and} \quad
\phi_4 \equiv \eta = \left( \begin{array}{c}
\eta^+ \\ \eta^0 \end{array} \right).
\ee
The doublet $\phi_1$ gives mass to the electron,
$\phi_2$ gives mass to the muon,
and $\phi_3$ gives mass to the $\tau$ lepton;
the doublet $\phi_4 \equiv \eta$ belongs to the dark sector.
We shall also use the conjugate doublets
$\tilde \phi_j = \left( \begin{array}{c}
{\phi_j^0}^\ast,\ - \phi_j^- \end{array} \right)^T$
and $\tilde \eta = \left( \begin{array}{c}
{\eta^0}^\ast,\ - \eta^- \end{array} \right)^T$.

The symmetries of our model are the following:
\begin{itemize}
\item
$\zz^\mathrm{(dark)}_2 \!\!\! :\ \eta \to - \eta,\
\nu_{eR} \to - \nu_{eR},\
\nu_{\mu R} \to - \nu_{\mu R}$,
and $\nu_{\tau R} \to - \nu_{\tau R}$.
This is an \emph{exact}\/ symmetry that prevents dark matter
from mixing with ordinary matter.
It is broken neither softly nor spontaneously,
because the vacuum expectation value (VEV) of $\eta$ is
zero.\footnote{Such scalar doublets have been dubbed `inert'
in ref.~\cite{barbieri}.}
\item The flavour lepton numbers $L_\alpha$.
They are broken \emph{only softly}\/
by the Majorana mass terms of the $\nu_{\alpha R}$:
\be
\label{majorana}
\mathcal{L}_\mathrm{Majorana} = - \frac{1}{2}
\left( \begin{array}{ccc}
\overline{\nu_{eR}}, & \overline{\nu_{\mu R}}, & \overline{\nu_{\tau R}}
\end{array} \right)
M_R
C \left( \begin{array}{c}
\overline{\nu_{eR}}^T \\ \overline{\nu_{\mu R}}^T \\ \overline{\nu_{\tau R}}^T
\end{array} \right)
+ \mathrm{H.c.},
\ee
where $C$ is the charge-conjugation matrix in Dirac space
and $M_R$ is a symmetric matrix in flavour space.
\item
$\zz_2^{(1)} \!\! :\ \phi_1 \to - \phi_1,\ e_R \to - e_R$,
$\zz_2^{(2)} \!\! :\ \phi_2 \to - \phi_2,\ \mu_R \to - \mu_R$,
and $\zz_2^{(3)} \!\! :\ \phi_3 \to - \phi_3,\ \tau_R \to - \tau_R$.
Because of these symmetries and of the $L_\alpha$,
the lepton Yukawa Lagrangian is
\bs
\label{yukawa}
\ba
\label{bnipy}
\mathcal{L}_{\ell\, \mathrm{Yukawa}} &=&
- y_1\, \overline{\nu_{eR}}\, \tilde \eta^\dagger D_{eL}
- y_2\, \overline{\nu_{\mu R}}\, \tilde \eta^\dagger D_{\mu L}
- y_3\, \overline{\nu_{\tau R}}\, \tilde \eta^\dagger D_{\tau L}
\\ & &
\label{bopyu}
- y_4\, \overline{e_R}\, \phi_1^\dagger D_{eL}
- y_5\, \overline{\mu_R}\, \phi_2^\dagger D_{\mu L}
- y_6\, \overline{\tau_R}\, \phi_3^\dagger D_{\tau L}
+ \mathrm{H.c.}
\ea
\es
The $\zz_2^{(j)}\ (j = 1, 2, 3)$ are broken spontaneously,
through the VEVs $\left\langle 0 \left| \phi_j^0 \right| 0 \right\rangle
= v_j \left/ \sqrt{2} \right.$,
to give mass to the charged leptons:
\be
\label{ghuop}
m_e = \left| \frac{y_4 v_1}{\sqrt{2}} \right|, \quad
m_\mu = \left| \frac{y_5 v_2}{\sqrt{2}} \right|, \quad
m_\tau = \left| \frac{y_6 v_3}{\sqrt{2}} \right|.
\ee
Besides,
the $\zz_2^{(j)}$ are also broken softly\footnote{ We recall
that in a renormalizable theory a symmetry is said to be broken softly
when all the symmetry-breaking terms have dimension smaller than four.
This leaves open two possibilities:
either they have both dimension two and dimension three
or they have only dimension two.
Soft symmetry breaking is consistent in quantum field-theoretic terms because,
when using it,
the dimension-four symmetry-violating terms generated by loops
are \emph{finite}.
The soft breaking of (super)symmetries is extensively used in model-building;
in particular,
all supersymmetric models contain soft supersymmetry-breaking terms.}
through quadratic terms in the scalar potential.
\item The $CP$ symmetry
\be
\label{cp}
CP:\ \left\{
\begin{array}{rcl}
D_L &\to& i \gamma_0 C\, S\, \overline{D_L}^T,
\\
\ell_R &\to& i \gamma_0 C\, S\, \overline{\ell_R}^T,
\\
\nu_R &\to& i \gamma_0 C\, S\, \overline{\nu_R}^T,
\\
\phi &\to& S\, \phi^\ast,
\\
\eta &\to& \eta^*,
\end{array} \right.
\quad \mbox{where}\
\begin{array}{l}
D_L = \left( \begin{array}{c} D_{eL} \\ D_{\mu L} \\ D_{\tau L} \end{array} \right),
\quad
\ell_R = \left( \begin{array}{c} e_R \\ \mu_R \\ \tau_R \end{array} \right),
\\*[6mm]
\nu_R = \left( \begin{array}{c}
\nu_{eR} \\ \nu_{\mu R} \\ \nu_{\tau R} \end{array} \right),
\quad
\phi = \left( \begin{array}{c} \phi_1 \\ \phi_2 \\ \phi_3 \end{array} \right).
\end{array}
\ee
Because of this symmetry,
in equation~(\ref{majorana})
\be
M_R = \left( \begin{array}{ccc}
x & y & y^\ast \\ y & z & w \\ y^\ast & w & z^\ast
\end{array} \right),
\ee
with \emph{real} $x$ and $w$,
\textit{i.e.}\ $S M_R S = M_R^\ast$;
moreover,
in equation~(\ref{yukawa}) $y_1$ and $y_4$ are real
and $y_3 = y_2^\ast,\ y_6 = y_5^\ast$.
Therefore,
\be
\frac{m_\mu}{m_\tau} = \left| \frac{v_2}{v_3} \right|,
\label{eq:mutau}
\ee
\textit{i.e.}\ the small ratio of muon to $\tau$-lepton mass
is explained through a small ratio of VEVs~\cite{small_ratio}.
The symmetry $CP$ is \emph{not broken softly}\footnote{We might accept
the soft breaking of $CP$ by quadratic terms in the scalar potential;
that soft breaking by terms of dimension two
would not disturb the dimension-three terms in $\mathcal{L}_\mathrm{Majorana}$.
But,
for the sake of simplicity,
we shall refrain in this paper from such a soft breaking.}
but it is broken spontaneously through the VEVs $v_j$,
especially through
$\left| v_2 \right| \neq \left| v_3 \right|$.\footnote{Ours is a model
of `real $CP$ violation',
\textit{i.e.}\ $CP$ violation originates in the inequality of two VEVs,
even if those VEVs are real~\cite{real}.}
\end{itemize}

As compared to the model in ref.~\cite{GL2003},
the present model has an extra doublet $\eta$,
whose vanishing VEV causes neutrino mass generation
to occur only at the one-loop level.
However,
as we will show below,
the very same mechanism that produces co-bimaximal mixing at the tree level
in the model of ref.~\cite{GL2003}
is effective at the one-loop level in the model of this paper.

In our model,
just as in the original model of Ma~\cite{ma2006},
dark matter may be either spin-one half---the lightest particle
arising from the mixture of $\nu_{eR}$,
$\nu_{\mu R}$,
and $\nu_{\tau R}$---or spin-zero---the lightest
of the two components $\varphi_{1,2}$ of $\eta^0$---depending on which of them
is lighter.
No other fields are needed in principle to account for the dark matter.

In the scalar potential,
a crucial role is played by the $CP$-invariant terms
\be
\label{pot}
\xi_1 \left[ \left( \phi_1^\dagger \eta \right)^2 +
\left( \eta^\dagger \phi_1 \right)^2 \right]
+ \xi_2 \left[ \left( \phi_2^\dagger \eta \right)^2 +
\left( \eta^\dagger \phi_3\right)^2 \right]
+ \xi_3 \left[ \left( \phi_3^\dagger \eta \right)^2 +
\left( \eta^\dagger \phi_2 \right)^2 \right],
\ee
where $\xi_1 = \xi_1^\ast$ and $\xi_3 = \xi_2^\ast$ because of Hermiticity.
Let us write
\be
\label{etgit}
\phi_4^0 \equiv \eta^0 = e^{i\gamma}\, \frac{\varphi_1 + i \varphi_2}{\sqrt{2}},
\ee
where the fields $\varphi_1$ and $\varphi_2$ are real
and the phase $\gamma$ is defined such that
\be
\mu^2 \equiv e^{2 i \gamma} \sum_{j = 1}^3 \xi_j\, \frac{ {v_j^\ast}^2}{2}
\ee
is real and positive.
Then,
the terms~(\ref{pot}) generate a mass term
\be
\label{fjodt}
\mu^2 \left( \varphi_1^2 - \varphi_2^2 \right),
\ee
which means that $\varphi_1$ and $\varphi_2$ are mass eigenfields
with distinct masses.
The term~(\ref{fjodt}) is the only one
that makes the masses of $\varphi_1$ and $\varphi_2$ different;
all other terms in the scalar potential contain $\left| \eta^0 \right|^2 =
\left. \left( \varphi_1^2 + \varphi_2^2 \right)\, \right/ 2$.

Now we make use of the results in appendix~A.
In the notation of equation~(\ref{Yukawa}),
equation~(\ref{bnipy}) means that $\Delta_1 = \Delta_2 = \Delta_3 = 0$
and $\Delta_4 = \diag \left( y_1, y_2, y_2^\ast \right)$;
notice that $S \Delta_4 S = \Delta_4^\ast$.
In the notation of equation~(\ref{gisfy}),
equation~(\ref{etgit}) reads $\mathcal{V}_{4 \varphi_1} = e^{i \gamma}$
and $\mathcal{V}_{4 \varphi_2} = i e^{i \gamma}$.
Then,
according to equation~(\ref{hopyi}),
$\Delta_{\varphi_1} = e^{i \gamma} \Delta_4$
and $\Delta_{\varphi_2} = i e^{i \gamma} \Delta_4$.
Applying equation~(\ref{final1}) we find the one-loop contribution to $\mnu$:
\be
\label{dML}
\delta \mnu = \frac{e^{2 i \gamma}}{32\pi^2} \left[ \Delta_4 W^\ast
\left( \frac{m_{\varphi_1}^2}{\widetilde m}
\ln{\frac{\widetilde m^2}{m_{\varphi_1}^2}} \right) W^\dagger \Delta_4 -
\Delta_4 W^\ast \left( \frac{m_{\varphi_2}^2 }{\widetilde m}
\ln{\frac{\widetilde m^2}{m_{\varphi_2}^2}} \right) W^\dagger \Delta_4 \right],
\ee
where the matrices $W$ and $\widetilde m$ are defined
through equation~(\ref{uidto}).
Note that there is no contribution to $\delta \mnu$ from a loop with $Z^0$
because the VEV of $\eta$ is assumed to vanish;
therefore,
the Dirac neutrino mass matrix $M_D$ in line~(\ref{woypf}) also vanishes.

In the limit $\mu^2 \to 0$,
the masses of $\varphi_1$ and $\varphi_2$ become equal
and the contributions of $\varphi_1$ and $\varphi_2$ to $\delta \mnu$
exactly cancel each other;
the light neutrinos then remain massless at the one-loop level~\cite{ma2006}.
This happens in the limit where all the terms in equation~(\ref{pot}) vanish.
Indeed,
in that limit the full Lagrangian is invariant under the $U(1)$ symmetry
\be
\label{gopyu}
D_L \to e^{i \psi} D_L, \quad \ell_R \to e^{i \psi} \ell_R,
\quad \eta \to e^{- i \psi} \eta,
\ee
which forbids light-neutrino masses~\cite{ma2006}.
We remark that there are,
in the scotogenic model of this paper,
several mechanisms for potentially suppressing the light-neutrino masses,
\textit{viz.}
\begin{itemize}
\item a large seesaw scale,
\textit{i.e.}\ large heavy-neutrino masses in $\widetilde m$;
\item
 small Yukawa couplings of $\nu_R$, \textit{i.e.}\ small $\Delta_4$;
\item small couplings $\xi_{1,2,3}$ in equation~(\ref{pot}),
hence $m_{\varphi_1}$ and $m_{\varphi_2}$ very close to each other,
because of an approximate symmetry~(\ref{gopyu});
\item the $\left( 32 \pi^2 \right)^{-1}$ factor in equation~\eqref{dML}
from the loop integral.
\end{itemize}
Let us present a benchmark
for all these suppressing factors.
Let both $\xi_{1,2,3}$ and $y_{1,2}$ be of order $10^{-2}$.
With $\left| v_{1,2,3} \right| \sim 100\, \mathrm{GeV}$
one then obtains
$\left| m_{\varphi_1} - m_{\varphi_2} \right| \sim 10\, \mathrm{GeV}$.
Assuming $m_{\varphi_{1,2}} \sim 100\, \mathrm{GeV}$,
one requires $\tilde m \sim 10^{7\mbox{--}8}\, \mathrm{GeV}$
in order to obtain $\delta \mnu \sim 0.1\, \mathrm{eV}$.
One concludes that the main suppression still originates
in the high seesaw scale.
However,
with small $\xi_{1,2,3}$ and $y_{1,2}$,
of order $10^{-3}$ or $10^{-4}$,
the seesaw scale could easily be in the TeV range
and thus accessible to the LHC.

Next we exploit the $CP$-invariance properties,
\textit{viz.}\ $S \Delta_4 S = \Delta_4^\ast$ and $S M_R S = M_R^\ast$.
Equation~(\ref{dML}) may be rewritten
\be
e^{-2i\gamma} \delta \mnu = \Delta_4 W^\ast \hat{A} W^\dagger \Delta_4,
\quad \mbox{with}\
\hat{A} = \frac{1}{32\pi^2} \left( \frac{m_{\varphi_1}^2}{\widetilde m}
\ln{\frac{\widetilde m^2}{m_{\varphi_1}^2}} -
\frac{m_{\varphi_2}^2 }{\widetilde m}
\ln{\frac{\widetilde m^2}{m_{\varphi_2}^2}} \right).
\ee
Now,
\be
\label{cp1}
S M_R S = M_R^\ast\ \Rightarrow \
\left( W^\dagger S W^\ast \right)^\ast \widetilde m
= \widetilde m \left( W^\dagger S W^\ast \right)\
\Rightarrow \
W^\dagger S W^\ast = X,
\ee
where $X$ is a diagonal sign matrix~\cite{GL2003}.
This is because,
according to the assumptions of the seesaw mechanism,
all the diagonal matrix elements of $\widetilde m$,
\textit{i.e.}\ all the heavy-neutrino masses,
are nonzero.
Using equation~(\ref{cp1}) we derive
\ba
S \left( \Delta_4 W^\ast \hat{A} W^\dagger \Delta_4 \right) S &=&
\left( S \Delta_4 S \right) \left( S W^\ast \right) \hat{A}
\left( W^\dagger S \right)
\left( S \Delta_4 S \right)
\no &=& \Delta_4^* W X \hat A X W^T \Delta_4^*
\no &=& \Delta_4^* W \hat A W^T \Delta_4^*
\no &=& \left( \Delta_4 W^* \hat A W^\dagger \Delta_4 \right)^*,
\ea
\textit{i.e.}
\be
S \left( e^{- 2 i \gamma} \delta \mnu \right) S =
\left( e^{- 2 i \gamma} \delta \mnu \right)^\ast.
\ee
Thus,
after a physically meaningless rephasing,
$\delta \mnu$ displays the defining feature~\eqref{sms} of co-bimaximal mixing.

\subsection{Approximation to the Higgs boson}

We use the notation of equation~(\ref{gisfy}).
The matrix $\mathcal{V}$ is complex $4 \times 8$ and,
according to ref.~\cite{osland},
\be
\tilde{\mathcal{V}}
= \left( \begin{array}{c}
\Re{\mathcal{V}} \\ \Im{\mathcal{V}} \end{array} \right)
\ee
is $8 \times 8$ orthogonal.
The last row of $\mathcal{V}$ corresponds to $\phi_4^0 \equiv \eta^0$.
For definiteness,
we let the last two columns of $\mathcal{V}$
correspond to $\varphi_1$ and $\varphi_2$,
which belong to the dark sector and do not mix with all the other scalars.
Therefore,
for practical purposes $\mathcal{V}$ is just a $3 \times 6$ matrix.
By definition,
$S^0_1 = G^0$ is the Goldstone boson and~\cite{osland}
\be
\label{v}
\mathcal{V}_{j1} = i\, \frac{v_j}{v} \quad \mbox{for}\ j = 1, 2, 3,
\quad \mbox{where}\
v \equiv \sqrt{\sum_{j=1}^3 \left| v_j \right|^2}.
\ee

The couplings of $S^0_b$ ($b = 2, \ldots, 6$)
to the gauge bosons are given by~\cite{osland}
\be
\label{bxiyp}
\frac{g}{v} \left( m_W W_\mu^+ W^{\mu -} + \frac{m_Z Z_\mu Z^\mu}{2 c_w} \right)
\sum_{b=2}^8 S^0_b\, \Re{\left( \sum_{j=1}^3 v_j^\ast \mathcal{V}_{jb} \right)}.
\ee
Therefore,
a given $S^0_b$ couples to the gauge bosons
with exactly the same strength as the Higgs boson
of the Standard Model if
\be
\label{vjorf}
g_{SVV} \equiv
\Re{\left( \sum_{j=1}^3 \frac{v_j^\ast}{v}\, \mathcal{V}_{jb} \right)} = 1.
\ee
Notice that,
because both three-vectors
$\left( v_1^\ast / v,\ v_2^\ast/v,\ v_3^\ast / v \right)$
and $\left( \mathcal{V}_{1b},\ \mathcal{V}_{2b},\ \mathcal{V}_{3b} \right)$
have unit modulus,
$\left| g_{SVV} \right| \le 1$.
Therefore,
equation~(\ref{vjorf}) holds in a limit situation.

According to equation~(\ref{bopyu}),
the scalars $S^0_b$ couple to the $\tau$ lepton through
\be
\label{gipst}
- \frac{1}{\sqrt{2}}\, \sum_{b=1}^6 S^0_b\, \overline{\tau}
\left( \mathcal{V}_{3b} y_6^\ast \gamma_R
+ \mathcal{V}_{3b}^\ast y_6 \gamma_L \right) \tau
=
- m_\tau \sum_{b=1}^6 S^0_b\, \overline{\tau}
\left( \frac{\mathcal{V}_{3b}}{v_3}\, \gamma_R
+ \frac{\mathcal{V}_{3b}^\ast}{v_3^\ast}\,
\gamma_L \right) \tau,
\ee
where $\gamma_{R,L}$ are the projectors of chirality in Dirac space.
In equation~(\ref{gipst}) we have assumed,
without loss of generality,
$y_6 v_3^*$ to be real and positive.
Therefore,
a given $S^0_b$ couples to the $\tau$ lepton in the same way
as the Higgs boson if
\be
\label{vutyd}
\frac{v \mathcal{V}_{3b}}{v_3} = 1.
\ee

\section{Extension of the model to the quark sector}
\label{quarks}

It is non-trivial to extend our model to the quark sector because
the $CP$ symmetry relates the Yukawa couplings of $\phi_2$ to those of $\phi_3$;
moreover,
some quarks must couple to $\phi_2$---and,
correspondingly,
other quarks must couple to $\phi_3$---in order that $CP$ violation,
which is generated through $v_1^\ast v_2 \neq v_1 v_3^\ast$,
manifests itself in the CKM matrix $V$.

We firstly expound some notation.
The quark Yukawa Lagrangian is
\be
\mathcal{L}_\mathrm{quark\, Yukawa} =
- \left( \begin{array}{ccc}
\overline{Q_{L1}},\ \overline{Q_{L2}},\ \overline{Q_{L3}}
\end{array} \right) \sum_{j=1}^3 \left[
\phi_j \Gamma_j
\left( \begin{array}{c} n_{R1} \\ n_{R2} \\ n_{R3} \end{array} \right)
+ \tilde \phi_j \Delta_j
\left( \begin{array}{c} p_{R1} \\ p_{R2} \\ p_{R3} \end{array} \right)
\right] + \mathrm{H.c.},
\ee
where $\overline{Q_{Lj}} = \left( \begin{array}{cc}
\overline{p_{Lj}}, & \overline{n_{Lj}} \end{array} \right)\, \mbox{for}\
j = 1, 2, 3$.
The mass matrices are
$M_n = \sum_{j=1}^3
\left( v_j \left/ \sqrt{2} \right. \right) \Gamma_j$
and $M_p = \sum_{j=1}^3
\left( v_j^\ast \left/ \sqrt{2} \right. \right) \Delta_j$.
They are diagonalized as
\bs
\label{vopry}
\ba
{U_L^n}^\dagger M_n U_R^n &=& \diag \left( m_d, m_s, m_b \right) \equiv M_d,
\\
{U_L^p}^\dagger M_p U_R^p &=& \diag \left( m_u, m_c, m_t \right) \equiv M_u,
\ea
\es
where the matrices $U_{L,R}^{n,p}$ are unitary.
The physical quarks are given by
\be
\left( \begin{array}{c} n_{R1} \\ n_{R2} \\ n_{R3} \end{array} \right)
= U_R^n
\left( \begin{array}{c} d_R \\ s_R \\ b_R \end{array} \right),
\quad
\left( \begin{array}{c} p_{R1} \\ p_{R2} \\ p_{R3} \end{array} \right)
= U_R^p
\left( \begin{array}{c} u_R \\ c_R \\ t_R \end{array} \right),
\ee
and analogously for the left-handed fields.
The quark mixing matrix is $V = {U_L^p}^\dagger U_L^n$.

\subsection{Extension 1}

One may include the quarks in the symmetries $\zz_2^{(j)}$ as follows:
\bs
\ba
\zz_2^{(1)}: & & \phi_1,\ e_R,\ \mbox{and}\ Q_{L1}\ \mathrm{change\ sign};
\\
\zz_2^{(2)}: & & \phi_2,\ \mu_R,\ \mbox{and}\ Q_{L2}\ \mathrm{change\ sign};
\\
\zz_2^{(3)}: & & \phi_3,\ \tau_R,\ \mbox{and}\ Q_{L3}\ \mathrm{change\ sign}.
\ea
\es
Then,
\bs
\ba
& &
\Gamma_1 = \left( \begin{array}{c} R_1 \\ 0_{1 \times 3} \\ 0_{1 \times 3}
\end{array} \right),
\quad
\Gamma_2 = \left( \begin{array}{c} 0_{1 \times 3} \\ R_2 \\ 0_{1 \times 3}
\end{array} \right),
\quad
\Gamma_3 = \left( \begin{array}{c} 0_{1 \times 3} \\ 0_{1 \times 3} \\ R_3
\end{array} \right),
\\*[1mm]
& &
\Delta_1 = \left( \begin{array}{c} R_1^\prime \\ 0_{1 \times 3} \\ 0_{1 \times 3}
\end{array} \right),
\quad
\Delta_2 = \left( \begin{array}{c} 0_{1 \times 3} \\ R_2^\prime \\ 0_{1 \times 3}
\end{array} \right),
\quad
\Delta_3 = \left( \begin{array}{c} 0_{1 \times 3} \\ 0_{1 \times 3} \\ R_3^\prime
\end{array} \right),
\ea
\es
where $R_{1,2,3}$ and $R^\prime_{1,2,3}$ are $1 \times 3$ row matrices.
Notice that both $\Gamma_{1,2,3}$ and $\Delta_{1,2,3}$
are in this extension rank~1 matrices.
The quark mass matrices are
\be
\label{vpowe}
M_n = \frac{1}{\sqrt{2}} \left( \begin{array}{c} R_1 v_1 \\ R_2 v_2 \\ R_3 v_3
\end{array} \right),
\quad
M_p = \frac{1}{\sqrt{2}}
\left( \begin{array}{c} R^\prime_1 v_1^\ast \\ R^\prime_2 v_2^\ast \\
R^\prime_3 v_3^\ast
\end{array} \right).
\ee
We define the $1 \times 3$ matrices
\be
\label{buiop}
\bar R_j \equiv R_j U_R^n, \quad \bar R^\prime_j \equiv R^\prime_j U_R^p.
\ee
Writing
\be
U_L^n = \left( \begin{array}{c} \hat R_1 \\ \hat R_2 \\ \hat R_3
\end{array} \right),
\quad
U_L^p = \left( \begin{array}{c} \hat R^\prime_1 \\ \hat R^\prime_2 \\
\hat R^\prime_3
\end{array} \right),
\ee
where the $\hat R_j$ and $\hat R^\prime_j$ are $1 \times 3$
row matrices,\footnote{Notice that the quark mixing matrix is
$V = \sum_{j=1}^3 \hat R^{\prime \dagger}_j \hat R_j$.}
one has,
from equations~(\ref{vopry}),
(\ref{vpowe}),
and~(\ref{buiop}),
\be
\bar R_j\, \frac{v_j}{\sqrt{2}} = \hat R_j M_d, \quad
\bar R^\prime_j\, \frac{v_j^\ast}{\sqrt{2}} = \hat R^\prime_j M_u.
\ee
(We do not use the summation convention.)
The Yukawa couplings
of the neutral scalars---see equation~(\ref{gisfy})--- are
\ba
& &
- \frac{1}{\sqrt{2}} \sum_{b=1}^8 S^0_b \sum_{j=1}^3 \left[
\overline{n_{Lj}}\, \mathcal{V}_{jb} R_j
\left( \begin{array}{c} n_{R1} \\ n_{R2} \\ n_{R3} \end{array} \right)
+ \overline{p_{Lj}}\, \mathcal{V}_{jb}^\ast R^\prime_j
\left( \begin{array}{c} p_{R1} \\ p_{R2} \\ p_{R3} \end{array} \right)
\right]
\no &=&
- \frac{1}{\sqrt{2}} \sum_{b=1}^8 S^0_b \sum_{j=1}^3 \left[
\overline{n_{Lj}}\, \mathcal{V}_{jb} \bar R_j
\left( \begin{array}{c} d_R \\ s_R \\ b_R \end{array} \right)
+ \overline{p_{Lj}}\, \mathcal{V}_{jb}^\ast \bar R^\prime_j
\left( \begin{array}{c} u_R \\ c_R \\ t_R \end{array} \right)
\right]
\no &=&
- \sum_{b=1}^8 S^0_b \sum_{j=1}^3 \left[
\overline{n_{Lj}}\, \frac{\mathcal{V}_{jb}}{v_j}\, \hat R_j M_d
\left( \begin{array}{c} d_R \\ s_R \\ b_R \end{array} \right)
+ \overline{p_{Lj}}\, \frac{\mathcal{V}_{jb}^\ast}{v_j^\ast}\, \hat R^\prime_j M_u
\left( \begin{array}{c} u_R \\ c_R \\ t_R \end{array} \right)
\right]
\no &=&
- \sum_{b=1}^8 S^0_b \left[
\left( \begin{array}{ccc} \overline{d_L}, & \overline{s_L}, & \overline{b_L}
\end{array} \right)
\sum_{j=1}^3 \frac{\mathcal{V}_{jb}}{v_j}\, \hat R_j^\dagger \hat R_j M_d
\left( \begin{array}{c} d_R \\ s_R \\ b_R \end{array} \right)
\right. \no & & \left.
+ \left( \begin{array}{ccc} \overline{u_L}, & \overline{c_L}, & \overline{t_L}
\end{array} \right)
\sum_{j=1}^3 \frac{\mathcal{V}_{jb}^\ast}{v_j^\ast}\,
\hat R^{\prime \dagger}_j \hat R^\prime_j M_u
\left( \begin{array}{c} u_R \\ c_R \\ t_R \end{array} \right)
\right].
\ea
Defining the Hermitian matrices
\be
H_j \equiv \hat R_j^\dagger \hat R_j =
\frac{\left| v_j \right|^2}{2}\, M_d^{-1} \bar R_j^\dagger \bar R_j M_d^{-1},
\quad
H^\prime_j \equiv \hat R_j^{\prime \dagger} \hat R_j^\prime =
\frac{\left| v_j \right|^2}{2}\,
M_u^{-1} \bar R_j^{\prime \dagger} \bar R_j^\prime M_u^{-1},
\ee
the Yukawa couplings of a given $S^0_b$ to the third-generation quarks
are given by
\be
- S^0_b \sum_{j=1}^3 \left[
m_b \left( H_j \right)_{33}
\overline{b} \left( \frac{\mathcal{V}_{jb}}{v_j}\, \gamma_R
+ \frac{\mathcal{V}_{jb}^\ast}{v_j^\ast}\, \gamma_L \right) b
+ m_t \left( H_j^\prime \right)_{33}
\overline{t} \left( \frac{\mathcal{V}_{jb}}{v_j}\, \gamma_L
+ \frac{\mathcal{V}_{jb}^\ast}{v_j^\ast}\, \gamma_R \right) t
\right].
\ee
Thus,
$S^0_b$ couples to the third-generation quarks
in the same way as the Higgs boson if
\be
\label{bvhit}
\sum_{j=1}^3 \left( H_j \right)_{33} \frac{\mathcal{V}_{jb}}{v_j}
= \sum_{j=1}^3 \left( H^\prime_j \right)_{33} \frac{\mathcal{V}_{jb}}{v_j}
= \frac{1}{v}.
\ee

We have not yet specified the way in which
the $CP$ symmetry is to be extended to the quark sector.
This may be chosen to be
\be
\label{vuidp}
CP: \quad \left\{ \begin{array}{rcl}
Q_L &\to& i \gamma_0 C\, S\, \overline{Q_L}^T, \\
n_R &\to& i \gamma_0 C\, \overline{n_R}^T, \\
p_R &\to& i \gamma_0 C\, \overline{p_R}^T,
\end{array} \right.
\quad \mbox{where} \quad
\begin{array}{lcl}
Q_L &=& \left( \begin{array}{c} Q_{L1},\ Q_{L2},\ Q_{L3}
\end{array} \right)^T,
\\*[1mm]
p_R &=& \left( \begin{array}{c} p_{R1},\ p_{R2},\ p_{R3}
\end{array} \right)^T,
\\*[1mm]
n_R &=& \left( \begin{array}{c} n_{R1},\ n_{R2},\ n_{R3}
\end{array} \right)^T.
\end{array}
\ee
The $CP$ symmetry~(\ref{vuidp}) enforces real $R_1$ and $R^\prime_1$ and
\be
R_3 = R_2^\ast, \quad R^\prime_3 = {R^\prime_2}^\ast.
\ee

\subsection{Extension 2}
\label{sec:ext2}

The extension of our model to the quark sector
expounded in the previous subsection
treats the down-type and up-type quarks in similar fashion.
It possesses flavour-changing
neutral Yukawa interactions (FCNYI) in both quark sectors.
In this subsection we suggest a different extension,
in which FCNYI are restricted to the up-type-quark sector.

Let the quarks be included in the symmetries $\zz_2^{(j)}$ as
\bs
\label{siutp}
\ba
\zz_2^{(1)}: & & \phi_1,\ e_R,\ p_{R1},\ n_{R1},\ n_{R2},\ \mbox{and}\ n_{R3}\
\mathrm{change\ sign};
\label{zz21} \\
\zz_2^{(2)}: & & \phi_2,\ \mu_R,\ \mbox{and}\ p_{R2}\ \mathrm{change\ sign};
\label{zz22} \\
\zz_2^{(3)}: & & \phi_3,\ \tau_R,\ \mbox{and}\ p_{R3}\ \mbox{change\ sign}.
\label{zz23}
\ea
\es
With this extension,
the Yukawa-coupling matrices $\Gamma_2$ and $\Gamma_3$ vanish outright.
In extension~2,
as distinct from extension~1,
the matrices $\Gamma_{2,3}$ are rank~0 while $\Gamma_1$ is rank~3.
Without loss of generality,
one may rotate the $Q_L$ and the $n_R$
so that $v_1 \Gamma_1 \left/ \sqrt{2} \right. = M_n$
is equal to $M_d$ from the outset.
Then,
$U_L^n = U_R^n = \bone_{3 \times 3}$ and the CKM matrix $V = {U_L^p}^\dagger$.

Analogously to equation~(\ref{gipst}),
the couplings of the neutral scalars to the down-type quarks
are given by
\be
- \sum_{b=1}^6 S^0_b \sum_{q = d, s, b} m_q\,
\overline{q} \left( \frac{\mathcal{V}_{1b}}{v_1}\, \gamma_R
+ \frac{\mathcal{V}_{1b}^\ast}{v_1^\ast}\, \gamma_L \right) q.
\ee
A given $S^0_b$ couples to the bottom quark
in the same way as the Higgs boson if
\be
\frac{v \mathcal{V}_{1b}}{v_1} = 1.
\ee

Now,
\bs
\label{eq:d}
\ba
\Delta_1 &\sim&
\left( \begin{array}{ccc} C_1, & 0_{3 \times 1}, & 0_{3 \times 1} \end{array} \right),
\label{eq:d1}
\\
\Delta_2 &\sim&
\left( \begin{array}{ccc} 0_{3 \times 1}, & C_2, & 0_{3 \times 1} \end{array} \right),
\\
\Delta_3 &\sim&
\left( \begin{array}{ccc} 0_{3 \times 1}, & 0_{3 \times 1}, & C_3 \end{array} \right),
\label{eq:d3}
\ea
\es
where $C_{1,2,3}$ are $3 \times 1$ column vectors.
The up-type-quark mass matrix is
\be
M_p = \frac{1}{\sqrt{2}}
\left( \begin{array}{ccc} C_1 v_1^\ast, & C_2 v_2^\ast, & C_3 v_3^\ast
\end{array} \right).
\label{eq:upmas}
\ee
We define the $3 \times 1$ column matrices $C^\prime_j \equiv {U_L^p}^\dagger C_j$.
Let
\be
U_R^p = \left( \begin{array}{c} R_1 \\ R_2 \\ R_3 \end{array} \right)
\ee
where the $R_{1,2,3}$ are $1 \times 3$ row matrices.
We know that
\be
{U_L^p}^\dagger M_p = \frac{1}{\sqrt{2}}
\left( \begin{array}{ccc}
C_1^\prime v_1^\ast, & C_2^\prime v_2^\ast, & C_3^\prime v_3^\ast
\end{array} \right)
= M_u {U_R^p}^\dagger
= \left( \begin{array}{ccc}
M_u R_1^\dagger, & M_u R_2^\dagger, & M_u R_3^\dagger
\end{array} \right).
\label{eq:UL}
\ee
Therefore,
$C^\prime_j = \left( \left. \sqrt{2} \right/ \! v_j^\ast \right) M_u R_j^\dagger$.

The couplings of the neutral scalars to the up-type quarks are given by
\ba
& & - \frac{1}{\sqrt{2}}
\sum_{b=1}^6 S^0_b\, \sum_{j=1}^3 \mathcal{V}_{jb}^\ast\
\overline{p_L}\, \Delta_j p_R\, + \mathrm{H.c.}
\no &=& - \frac{1}{\sqrt{2}}
\sum_{b=1}^6 S^0_b\, \sum_{j=1}^3 \mathcal{V}_{jb}^\ast\
\overline{p_L}\, C_j p_{Rj}\, + \mathrm{H.c.}
\no &=&
- \frac{1}{\sqrt{2}}
\sum_{b=1}^6 S^0_b\, \sum_{j=1}^3 \mathcal{V}_{jb}^\ast
\left( \begin{array}{ccc} \overline{u_L}, & \overline{c_L}, &
\overline{t_L} \end{array} \right)
{U_L^p}^\dagger C_j R_j
\left( \begin{array}{c} u_R \\ c_R \\ t_R \end{array} \right)
+ \mathrm{H.c.}
\no &=&
- \frac{1}{\sqrt{2}}
\sum_{b=1}^6 S^0_b\, \sum_{j=1}^3 \mathcal{V}_{jb}^\ast
\left( \begin{array}{ccc} \overline{u_L}, & \overline{c_L}, &
\overline{t_L} \end{array} \right)
C_j^\prime R_j
\left( \begin{array}{c} u_R \\ c_R \\ t_R \end{array} \right)
+ \mathrm{H.c.}
\no &=&
- \sum_{b=1}^6 S^0_b\, \sum_{j=1}^3 \frac{\mathcal{V}_{jb}^\ast}{v_j^\ast}
\left( \begin{array}{ccc} \overline{u_L}, & \overline{c_L}, &
\overline{t_L} \end{array} \right)
M_u R_j^\dagger R_j
\left( \begin{array}{c} u_R \\ c_R \\ t_R \end{array} \right)
+ \mathrm{H.c.}
\label{biutp}
\ea
Let us define $H_j = R_j^\dagger R_j$.
The $H_j$ are three Hermitian matrices;
since $U_R^p$ is unitary,
$H_1 + H_2 + H_3 = \bone_{3 \times 3}$.
The couplings of the neutral scalars to the top quark are given by
\be
- m_t \sum_{b=1}^6 S^0_b\, \sum_{j=1}^3
\left( H_j \right)_{33}
\overline{t}
\left(  \frac{\mathcal{V}_{jb}^\ast}{v_j^\ast}\, \gamma_R
+  \frac{\mathcal{V}_{jb}}{v_j} \gamma_L \right) t.
\ee
Thus,
one given scalar $S^0_b$ couples to the top quark
in the same way as the Higgs boson if
\be
\sum_{j=1}^3 \frac{v \mathcal{V}_{jb} \left( H_j \right)_{33}}{v_j} = 1.
\ee

For the action of $CP$ in the quark sector we choose
\be
CP: \quad \left\{ \begin{array}{rcl}
Q_L &\to& i \gamma_0 C\, \overline{Q_L}^T, \\
n_R &\to& i \gamma_0 C\, \overline{n_R}^T, \\
p_R &\to& i \gamma_0 C\, S\, \overline{p_R}^T.
\end{array} \right.
\ee
In this way,
\be
\label{buhio}
C_1 = C_1^\ast, \quad C_2 = C_3^\ast.
\ee

\section{The scalar potential}
\label{Higgs}

The scalar potential may be separated into three pieces:
\be
V = V_\eta + V_\mathrm{symmetric} + V_\mathrm{soft}.
\label{eq:V}
\ee
By definition,
all the terms containing $\eta$ belong to $V_\eta$,
whereas $V_\mathrm{symmetric} + V_\mathrm{soft}$ contains
exclusively the $\phi_j$ with $j = 1, 2, 3$.
By definition,
$V_\eta + V_\mathrm{symmetric}$
is invariant under all the symmetries of the Lagrangian,
\textit{i.e.}~under both $CP$ and the $\zz_2^{(j)}$ for $j = 1, 2, 3$,
whereas $V_\mathrm{soft}$ breaks the $\zz_2^{(j)}$ softly but preserves $CP$.
Obviously,
in any term in $V_\eta + V_\mathrm{symmetric}$
only even numbers of $\eta$'s and of each of the  $\phi_j$'s
can occur.
We have
\ba
V_\eta &=&
\mu_\eta\, \eta^\dagger \eta
+ \tilde \lambda_1 \left( \eta^\dagger \eta \right)^2
\no & &
+ \eta^\dagger \eta \left[ \tilde \lambda_2\, \phi_1^\dagger \phi_1
+ \tilde \lambda_3 \left( \phi_2^\dagger \phi_2 + \phi_3^\dagger \phi_3 \right)
\right]
+ \tilde \lambda_4\, \eta^\dagger \phi_1\, \phi_1^\dagger \eta
+ \tilde \lambda_5 \left( \eta^\dagger \phi_2\, \phi_2^\dagger \eta
+ \eta^\dagger \phi_3\, \phi_3^\dagger \eta \right)
\no & &
+ \xi_1 \left[
\left( \phi_1^\dagger \eta \right)^2
+ \left( \eta^\dagger \phi_1 \right)^2 \right]
+ \left\{ \xi_2 \left[ \left( \phi_2^\dagger \eta \right)^2
+ \left( \eta^\dagger \phi_3 \right)^2 \right]
+ \mathrm{H.c.} \right\},
\ea
where $\tilde \lambda_{1\mbox{--}5}$ and $\xi_1$ are real
while $\xi_2$ is in general complex.
We assume that the real coefficient $\mu_\eta$ is \emph{positive},
so that the VEV $\langle \eta^0 \rangle_0$ vanishes.
Moreover,
$\mu_\eta$ must be sufficiently larger than the Fermi scale squared,
so that the terms with coefficients $\tilde \lambda_{2\mbox{--}5}$
cannot make $\mu_\eta \to \mu_\eta
+ \left( \tilde \lambda_2 + \tilde \lambda_4 \right) \left| v_1 \right|^2
+ \left( \tilde \lambda_3 + \tilde \lambda_5 \right)
\left( \left| v_2 \right|^2 + \left| v_3 \right|^2 \right)$ become negative.

Due to the symmetries,
$V_\mathrm{symmetric}$ is given by
\ba
V_\mathrm{symmetric}&=&
\mu_1\, \phi_1^\dagger \phi_1 +
\mu_2 \left( \phi_2^\dagger \phi_2 + \phi_3^\dagger \phi_3 \right)
+ \lambda_1 \left( \phi_1^\dagger \phi_1 \right)^2
+ \lambda_2 \left[ \left( \phi_2^\dagger \phi_2 \right)^2 +
\left( \phi_3^\dagger \phi_3 \right)^2 \right]
\no & &
+ \lambda_3\, \phi_1^\dagger \phi_1
\left( \phi_2^\dagger \phi_2 + \phi_3^\dagger \phi_3 \right)
+ \lambda_4 \left( \phi_1^\dagger \phi_2\, \phi_2^\dagger \phi_1 +
\phi_1^\dagger \phi_3\, \phi_3^\dagger \phi_1 \right)
\no & &
+ \lambda_5\, \phi_2^\dagger \phi_2\, \phi_3^\dagger \phi_3
+ \lambda_6\, \phi_2^\dagger \phi_3\, \phi_3^\dagger \phi_2
+ \lambda_7 \left( \phi_2^\dagger \phi_3 \right)^2
+ \lambda_7^* \left( \phi_3^\dagger \phi_2 \right)^2
\no & &
+ \lambda_8 \left[ \left( \phi_1^\dagger \phi_2 \right)^2 +
\left( \phi_3^\dagger \phi_1 \right)^2 \right]
+ \lambda_8^* \left[ \left( \phi_2^\dagger \phi_1 \right)^2 +
\left( \phi_1^\dagger \phi_3 \right)^2 \right],
\label{eq:Vsim}
\ea
with real $\lambda_l$ for $l = 1, \ldots, 6$
and complex $\lambda_7$ and $\lambda_8$.
The soft-breaking potential,
which consists of terms of dimension two and abides by the $CP$ symmetry,
is
\be
V_\mathrm{soft} = \mu_3\,\phi_2^\dagger \phi_3 +
\mu_3^\ast\,\phi_3^\dagger \phi_2 +
\mu_4 \left( \phi_1^\dagger \phi_2 + \phi_3^\dagger \phi_1 \right) +
\mu_4^\ast \left( \phi_2^\dagger \phi_1 + \phi_1^\dagger \phi_3 \right).
\label{eq:Vsoft}
\ee

We firstly follow ref.~\cite{small_ratio} to investigate the minimum of
$V_\mathrm{symmetric}$.
We write $\lambda_l = \left| \lambda_l \right| e^{i \alpha_l}$ for $l = 7, 8$ and
\be
v_1 = w_1 e^{i \beta_1},
\quad
v_2 = w \sin{\sigma}\, e^{i \beta_2},
\quad
v_3 = w \cos{\sigma}.
\ee
We require without loss of generality that $w_1 \geq 0$,
$w \geq 0$,
and $\sigma$ is in the first quadrant.
Let $F$ denote the sum of the terms of
$\left\langle 0 \left| V_\mathrm{symmetric} \right| 0 \right\rangle$
that have a non-trival $\sigma$-dependence.
One has
\bs
\label{F}
\ba
4 F &=&
\lambda_2 w^4 \left( \sin^4 \sigma + \cos^4 \sigma \right)
+ \left( \lambda_5 + \lambda_6 \right) w^4 \sin^2 \sigma \cos^2 \sigma
\label{fipt} \\ & &
+ 2 \left| \lambda_7 \right| w^4 \sin^2{\sigma} \cos^2{\sigma}
\cos{\left( \alpha_7 - 2 \beta_2 \right)}
\label{hipy} \\ & &
+ 2 \left| \lambda_8 \right| w_1^2 w^2 \left[
\sin^2{\sigma} \cos{\left( \alpha_8 + 2 \beta_2 - 2 \beta_1 \right)}
+ \cos^2{\sigma} \cos{\left( \alpha_8 + 2 \beta_1 \right)}
\right].
\label{ghuiy}
\ea
\es
In line~(\ref{fipt}) we use
$\sin^4{\sigma} + \cos^4{\sigma} = 1 - 2 \sin^2{\sigma} \cos^2{\sigma}$.
We require
\be
- 2 \lambda_2 + \lambda_5 + \lambda_6 - 2 \left| \lambda_7 \right| > 0.
\ee
Then,
the minimum of lines~(\ref{fipt}) and~(\ref{hipy})
is achieved when $\sin{\sigma} \cos{\sigma} = 0$.
This may also be the minimum of line~(\ref{ghuiy}) because
\be
\sin^2{\sigma} \cos{\left( \alpha_8 + 2 \beta_2 - 2 \beta_1 \right)}
+ \cos^2{\sigma} \cos{\left( \alpha_8 + 2 \beta_1 \right)}
\geq -1
\ee
and the value $-1$
can always be obtained,
irrespective of the value of $\sigma$,
through suitable choices of $\beta_1$ and $\beta_2$.
Thus,
assuming $\sin{\sigma} = 0$ instead of $\cos{\sigma} = 0$,
the minimum of $F$ is at $\sigma = 0$,
\textit{i.e.}\ $v_2 = 0$,
and $\alpha_8 + 2 \beta_1 = \pi$.
The latter relation,
however,
is irrelevant if $v_1 = 0$,
because then $\beta_1$ is meaningless.
We assume,
indeed,
that the coefficient $\mu_1$ is positive and so large that $v_1 = 0$.
The minimum of $V_\mathrm{symmetric}$ then has $v_1 = v_2 = 0$.

In the limit $v_1 = v_2 = 0$ it is easy to compute the scalar mass spectrum.
Writing
\be
\phi_1^0 = e^{- i \alpha_8 / 2}\, \frac{\rho_1 + i\sigma_1}{\sqrt{2}},
\quad
\phi_2^0 = e^{i \alpha_7 / 2}\, \frac{\rho_2 + i\sigma_2}{\sqrt{2}},
\quad
\phi_3^0 = \frac{w + \rho_3 + i\sigma_3}{\sqrt{2}}
\ee
we find
\bs
\label{app}
\ba
m^2_{\phi_1^+} &=& \mu_1 + \frac{\lambda_3}{2}\, w^2,
\label{phi1}
\\
m^2_{\rho_1} &=& \mu_1 + \left( \frac{\lambda_3 + \lambda_4}{2}
+ \left| \lambda_8 \right| \right) w^2,
\label{rho1}
\\
m^2_{\sigma_1} &=& \mu_1 + \left( \frac{\lambda_3 + \lambda_4}{2}
- \left| \lambda_8 \right| \right) w^2,
\label{sigma1}
\\
m^2_{\phi_2^+} &=& \left( - \lambda_2 + \frac{\lambda_5}{2} \right) w^2,
\label{phi2}
\\
m^2_{\rho_2} &=& \left( - \lambda_2 + \frac{\lambda_5 + \lambda_6}{2}
+ \left| \lambda_7 \right| \right) w^2,
\label{rho2}
\\
m^2_{\sigma_2} &=& \left( - \lambda_2 + \frac{\lambda_5 + \lambda_6}{2}
- \left| \lambda_7 \right| \right) w^2,
\label{sigma2}
\\
m^2_{\rho_3} &=& 2 \lambda_2 w^2.
\label{rho3}
\ea
\es
Moreover,
$\sigma_3 = G^0$ is the neutral Goldstone boson
and $\phi_3^+$ is the charged Goldstone boson,
which are absorbed by the $Z^0$ and the $W^+$ gauge bosons,
respectively.
The fields $\phi_1^+$,
$\rho_1$,
and $\sigma_1$ are heavy because of the large $\mu_1$.
The scalar $\rho_3$ is to be identified with the Higgs boson.
From $m_{\rho_3} \simeq 125\, \mathrm{GeV}$ and $w \simeq 246\, \mathrm{GeV}$
one obtains $\lambda_2 \simeq 0.13$.
The masses of $\phi_2^+$,
$\rho_2$,
and $\sigma_2$ cannot be very large if one wants
to stay in the perturbative regime with respect to $\lambda_{5,6,7}$.

Now we take into account $V_\mathrm{soft}$.
This generates $v_{1,2} \neq 0$ due to the presence of terms linear
in $\phi_1$ and $\phi_2$.
For sufficiently small $v_1$ and $v_2$,
\bs
\ba
v_1 &\simeq& - \frac{\mu_4^\ast v_3}
{\mu_1 + \frac{1}{2} \left( \lambda_3 + \lambda_4 \right) \left| v_3 \right|^2
+ \lambda_8^\ast\, e^{-2i\beta_1} v_3^2},
\\
v_2 &\simeq& - \frac{\mu_3 v_3}
{\mu_2 + \frac{1}{2} \left( \lambda_5 + \lambda_6 \right)
\left| v_3 \right|^2 + \lambda_7\, e^{-2i\beta_1} v_3^2}.
\ea
\es

In order to verify whether the scalar potential of
equations~\eqref{eq:V}--\eqref{eq:Vsoft} can
produce a vacuum with the desired hierarchy of VEVs
and yield acceptable scalar masses and couplings,
we have performed a numerical scan of the parameter space of the potential.
We have taken all the parameters of the scalar potential to be real,
and we have also assumed real VEVs,
parameterized as
\be
v_1 = v \cos{\beta}, \quad
v_2 = \pm \frac{v\, m_\mu \sin{\beta}}{\sqrt{m_\mu^2 + m_\tau^2}}, \quad
v_3 = \frac{v\, m_\tau \sin{\beta}}{\sqrt{m_\mu^2 + m_\tau^2}},
\label{eq:vevs}
\ee
where the angle $\beta$ may be either in the first or second quadrant.
In this way we satisfy equation~\eqref{eq:mutau}.
It is desirable to have $\beta$ close to $\pi/2$
so that $\left| v_1 \right|$ is much smaller than $\left| v_{2,3} \right|$,
because $m_e \propto v_1$,
\textit{cf.}\ equation~\eqref{ghuop};
in our scan we have restricted
$5 \le \left| \tan{\beta} \right| \le 400$.

Notice that equation~\eqref{eq:vevs}
is used just as an \textit{Ansatz}\/ for our numerical study:
nothing guarantees that the \emph{global}\/ minimum of the potential
has real VEVs or,
indeed,
that it conserves the $U(1)$ of electromagnetism.
We also remind that,
since in our model the $CP$ transformation
effects $\phi_2 \leftrightarrow \phi_3^\ast$,
a vacuum with $v_2 \neq v_3$ will in general lead to $CP$ violation
even when the VEVs are real---indeed,
we shall use equation~\eqref{eq:vevs} to fit for the observed $CP$ violation,
\textit{cf.}\ equations~\eqref{vusyp} below.

We have made the quartic couplings of the potential
comply with certain basic restrictions for the model to make sense:
\begin{itemize}
\item The scalar potential has to be bounded from below (BFB),
\textit{i.e.}~there should be no directions in field space
along which the potential can tend to minus infinity.
To find the BFB conditions
one must study the behaviour of the scalar potential
for specific directions along which the fields may tend to infinity
and verify which combinations of parameters ensure that the potential is BFB.
The set of necessary conditions\footnote{A set
of necessary \emph{and sufficient}\/ BFB conditions
was obtained for the two-Higgs-doublet model in ref.~\cite{Ivanov:2006yq},
but the procedure described therein
cannot be generalized to models with a larger scalar content.}
that we have enforced is
(see refs.~\cite{Branco:2011iw,kannike})
\bs
\label{kkannike}
\ba
\lambda_1 &>& 0,
\\
\lambda_2 &>& 0,
\\
L_1 \equiv 2 \sqrt{\lambda_1\lambda_2} + \lambda_3
+ \left( \lambda_4 - 2 \left| \lambda_8 \right| \right)
\Theta \left( 2 \left| \lambda_8 \right| - \lambda_4 \right)
&>& 0,
\\
L_2 \equiv 2 \lambda_2 + \lambda_5
+ \left( \lambda_6 - 2 \left| \lambda_7 \right| \right)
\Theta \left( 2 \left| \lambda_7 \right| - \lambda_6 \right)
&>& 0,
\\
\sqrt{\lambda_1} L_2
+ 2 \sqrt{\lambda_2} L_1
- 4 \lambda_2 \sqrt{\lambda_1}
+ L_1 \sqrt{L_2} &>& 0,
\ea
\es
%
where $\Theta$ denotes the step function of Heaviside.
\item The model must respect unitarity and perturbativity.
Therefore,
the quartic couplings of the potential cannot be arbitrarily large.
We have imposed $\left| \lambda_l \right| \leq 20\ \forall l = 1, \ldots, 8$;
this should guarantee appropriate behaviour.
\item The model has to obey
the phenomenological constraint on the oblique parameter $T$,
\textit{viz.}~$T = 0.01 \pm 0.12$~\cite{Agashe:2014kda}.
The value of $T$ was computed through the formulae
of ref.~\cite{osland}.\footnote{We have explicitly checked that
the bounds on the oblique parameter $S$ never give
additional restrictions to this model.}
\end{itemize}

In the further discussion of this section,
we use the following notation for the exact scalar mass eigenstates:
charged scalars $H^+_i$ ($i=1,2$),
$CP$-even neutral scalars $h_i$ ($i=1,2,3$),
and pseudoscalars $A_i$
($i=1,2$).\footnote{Note that,
since in our fit we have assumed both the parameters of the potential
and the VEVs to be real,
the scalar sector of the model conserves $CP$,
hence there are well-defined neutral scalars and pseudoscalars.}
There is the correspondence $H^+_i \leftrightarrow \phi^+_i$,
$h_i \leftrightarrow \rho_i$,
and $A_i \leftrightarrow \sigma_i$
between the exact and the approximate mass eigenstates,
with the approximate
masses given by equations~(\ref{app}).
By definition,
the mass of $A_1$ is larger than the mass of $A_2$
and the mass of $H_1^+$ is larger than the mass of $H_2^+$;
similarly,
$m_{h_1} > m_{h_2} > m_{h_3}$.

With the above restrictions in place,
we still have to implement in our numerical scan a scalar state $h_3$,
corresponding to the Higgs boson,
with mass $125 \pm 1\, \mathrm{GeV}$ and almost ``aligned'',
according to the discussion held in the previous sections,
with the $\rho_3$ direction.
Also,
since the model does have FCNYI,
it is very convenient that all the neutral scalars
other than the lightest one be as heavy as possible.
In our scan we have imposed a lower bound of $600\, \mathrm{GeV}$
on the masses of the charged scalars $H^+_{1,2}$,
of the pseudoscalars $A_{1,2}$,
and of the heavier $CP$-even scalars $h_{1,2}$.\footnote{We have also imposed
an upper bound of $1500\, \mathrm{GeV}$ on all the scalar masses.}
Moreover,
since the experimental constraints on FCNYI in the down-type-quark sector
are much stronger than those in the up-type-quark sector,
we have chosen to scan exclusively the extension 2 of our model
to the quark sector,
since that extension has no FCNYI in the down sector.
Finally,
in order to comply with current LHC experimental results~\cite{LHC},
the 125\,GeV-mass scalar $h_3$
must have couplings to the gauge bosons
and to the heavy fermions close to the SM values.
Specifically,
in our scan we have demanded that:
\begin{itemize}
\item The coupling of $h_3$ to the gauge bosons be within 10\% of its
expected SM value,
\textit{i.e.}
\be
g_{SVV} =
\frac{v_1 \mathcal{V}_{13} + v_2 \mathcal{V}_{23} +
v_3 \mathcal{V}_{33}}{v} \ge 0.9,
\ee
with the matrix $\mathcal{V}$ defined in equation~\eqref{gisfy}.
By definition,
the third column of that matrix corresponds
to the $125\, \mathrm{GeV}$-mass neutral scalar $h_3$.
\item The coupling of $h_3$ to the bottom quarks be within 10\% of its
expected SM value,
\textit{i.e.}
\be
g_{Sbb} \equiv
\frac{v \mathcal{V}_{13}}{v_1} = 1 \pm 0.1.
\ee
\item The coupling of $h_3$ to the tau leptons be within 10\% of its
expected SM value,
\textit{i.e.}
\be
g_{S\tau\tau} \equiv
\frac{v \mathcal{V}_{33}}{v_3} = 1 \pm 0.1.
\ee
\end{itemize}

\begin{figure}[t]
\centerline{\epsfig{file=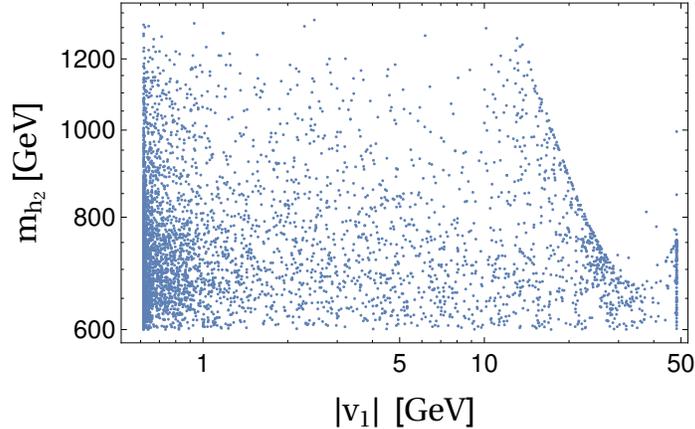,width=0.6\textwidth}}
\vspace*{-3mm}
\caption{Scatter plot of
the mass of the second heaviest $CP$-even scalar
\textit{versus}\/ the absolute value of $v_1$.}
\vspace*{3mm}
\label{fig:v1mh2}
\end{figure}

In figure~\ref{fig:v1mh2}
we have plotted the mass of the second heaviest $CP$-even scalar,
$h_2$,
against the value of the VEV $v_1$.
There are two features worth mentioning.
Firstly,
the value of $\left| v_1 \right|$ may be very small,
\textit{i.e.}\ the value of
$\left| \tan{\beta} \right|$
may be very large.
Secondly,
the mass of $h_2$
is never higher than $1.35\, \mathrm{TeV}$.
The first feature implies $\sin{\beta} \simeq 1$;
therefore,
the values of $v_2$ and $v_3$ in equations~\eqref{eq:vevs}
are essentially constant:
$\left| v_2 \right| \simeq 14.6$\,GeV and $v_3 \simeq 245.3$\,GeV.
The second feature arises
from the need to keep the magnitudes of the quartic couplings
in the perturbative regime,
\textit{viz.}~$\left| \lambda_i \right| \le 20$ for $i = 1, 3, 4, 5, 6$.
The other quartic couplings retain smaller magnitudes;
we obtained $-12 < \lambda_7 < 13$ and $ -14 < \lambda_8 < 2$ in our scan,
while the coupling $\lambda_2 \in \left[ 0.12,\ 0.14 \right]$
as predicted above.

\begin{figure}[t]
\centerline{\epsfig{file=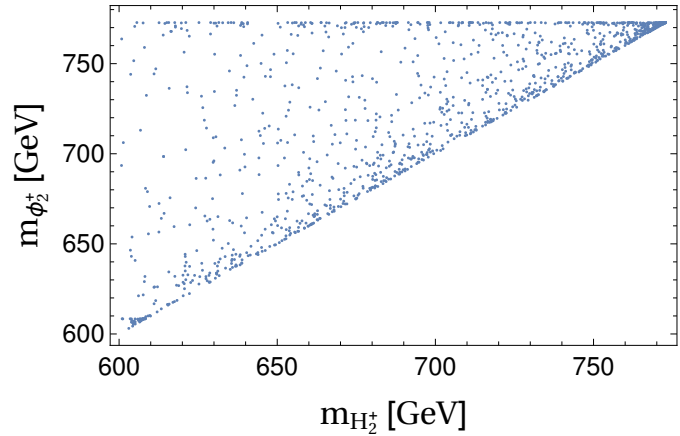,width=0.6\textwidth}}
\vspace*{-3mm}
\caption{Scatter plot of
the approximate expression~\eqref{phi2}
for the mass of the lightest charged scalar
\textit{versus}\/ the true mass of that particle.}
\vspace*{3mm}
\label{fig:mch2}
\end{figure}

Since we can find regions in the parameter space
for which $\left| v_1 \right| \ll \left| v_2 \right| \ll \left| v_3 \right|$,
the expressions~\eqref{app} constitute good approximations
to the exact scalar masses.
To illustrate this,
in figure~\ref{fig:mch2}
we have plotted the exact mass of $H^+_2$
against the approximate expression for that mass in equation~\eqref{phi2}.
As we can appreciate from the plot,
the approximate formula describes quite reasonably the true value,
though deviations $\lesssim 30\%$ occur in some cases.
Similar results have been obtained for the approximate formulae
for the masses of $h_2$,
$A_2$,
$h_1$,
$A_1$,
and $H^+_1$.

\begin{figure}[t]
\centerline{\epsfig{file=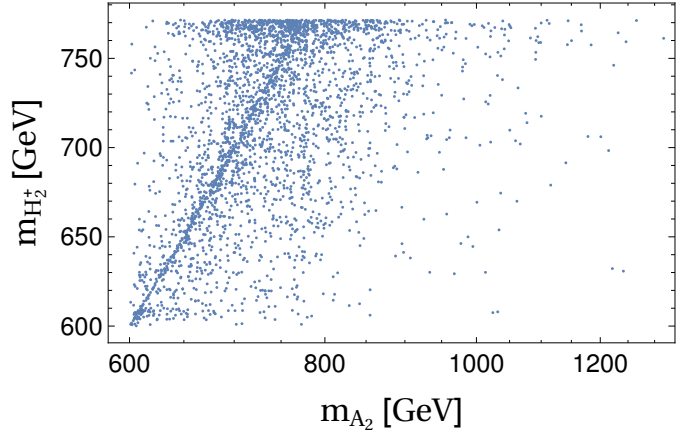,width=0.6\textwidth}}
\vspace*{-3mm}
\caption{Scatter plot of
the mass of the lightest charged scalar
\textit{versus}\/
the mass of the lightest pseudoscalar.}
\vspace*{3mm}
\label{A2H2}
\end{figure}

\begin{figure}[t]
\centerline{\epsfig{file=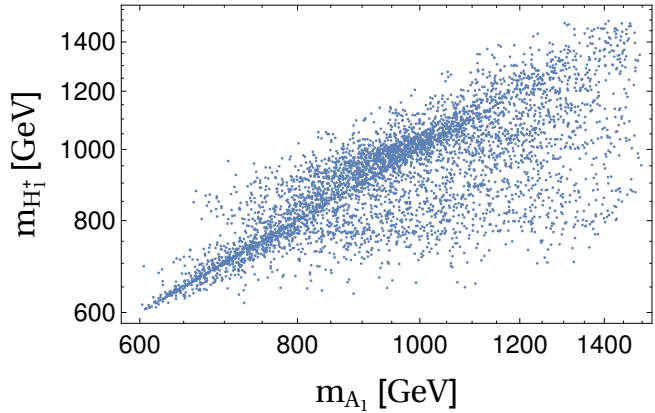,width=0.6\textwidth}}
\vspace*{-3mm}
\caption{Scatter plot of
the mass of the heaviest charged scalar
\textit{versus}\/
the mass of the heaviest pseudoscalar.}
\vspace*{3mm}
\label{A1H1}
\end{figure}

In figure~\ref{A2H2}
one observes that the mass of $A_2$ may be smaller than,
but may also be as much as twice, the one of $H_2^+$.
This is in spite of our enforcement of the experimental bound
on the oblique parameter $T$,
which might suggest the masses of $A_2$,
$h_2$,
and $H_2^+$ to be almost degenerate;
they are not.
In figure~\ref{A1H1}
one observes the same as in figure~\ref{A2H2},
but now for the heaviest charged scalar and the heaviest pseudoscalar.
One sees once again that the masses
of the heaviest scalars can differ considerably.

\begin{figure}[t]
\centerline{\epsfig{file=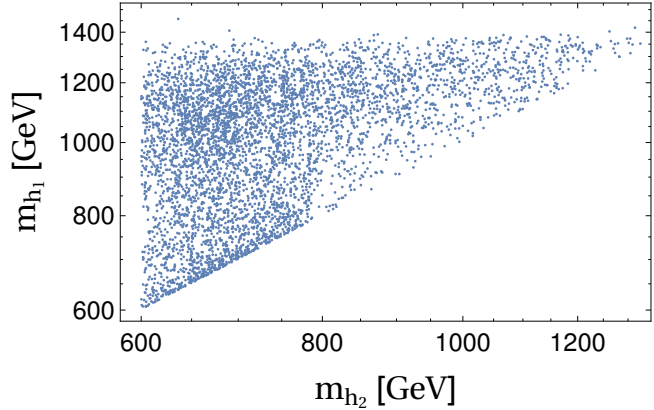,width=0.6\textwidth}}
\vspace*{-3mm}
\caption{Scatter plot of
the mass of the heaviest neutral scalar
\textit{versus}\/
the mass of the intermediate-mass neutral scalar.}
\vspace*{3mm}
\label{h1h2}
\end{figure}

Comparing figures~\ref{A2H2} and~\ref{A1H1} one sees that
the masses of the heaviest and the lightest scalars
are not necessarily much different.
This can be confirmed through figure~\ref{h1h2},
where the masses of the two heavy neutral scalars
are plotted against each other.
One sees the $m_{h_1}$ and $m_{h_2}$ may be quite close to each other,
whatever their average value.

At this stage,
we have shown that
our model can reproduce a boson with mass roughly $125\, \mathrm{GeV}$
and couplings to the gauge bosons,
to the bottom quarks,
and to the tau leptons close to the expected SM values.
The Yukawa-coupling matrix $\Gamma_1$,
as described in section~\ref{sec:ext2},
reproduces the known down-type-quark masses.
We now have to show that
the model can also reproduce both the up-type-quark masses and the CKM matrix.
We take the values of the up-type-quark running masses,
at the scale $m_Z$,
from ref.~\cite{Xing:2011aa}
and the values of the CKM-matrix parameters from ref.~\cite{Agashe:2014kda}:
\bs
\label{vusyp}
\ba
& &
m_u = 1.38^{+0.42}_{-0.41}\, \mathrm{MeV}, \quad
m_c = 638^{+43}_{-84}\, \mathrm{MeV}, \quad
m_t = 172.1 \pm 1.2\, \mathrm{GeV},
\\
& &
\left| V_{us} \right| = 0.22536 \pm 0.00061, \quad
\left| V_{cb} \right| = 0.04114 \pm 0.0012, \quad
\\
& &
\left| V_{ub} \right| = \left( 355 \pm 15 \right) \times 10^{-5}, \quad
J = 306^{+21}_{-20} \times 10^{-7}.
\ea
\es
We have used the values of the parameters of the model
that had previously been shown to obey all the constraints hitherto mentioned
and we have searched for phenomenologically acceptable values
for the Yukawa couplings in equations~\eqref{buhio}:
\be
C_1 = \left( \begin{array}{c} f_1 \\ f_2 \\ f_3 \end{array} \right),
\quad
C_2 = \left( \begin{array}{c} f_4 + i f_5 \\ f_6 + i f_7 \\ f_8 + i f_9
\end{array} \right),
\quad
C_3 = \left( \begin{array}{c} f_4 - i f_5 \\ f_6 - i f_7 \\ f_8 - i f_9
\end{array} \right),
\ee
with real $f_1,\ldots,f_9$.
Specifically,
we have demanded in our fitting procedure that
the up-type-quark mass matrix in equation~\eqref{eq:upmas},
after being diagonalized as $V M_p M_p^\dagger V^\dagger = \mathrm{diag}
\left( m_u^2, m_c^2, m_t^2 \right)$,
gives both the right values for the masses of the up-type quarks
and for the moduli of the matrix elements of the CKM matrix $V$.
We moreover require that the coupling of $h_3$ to top quarks
be within 10\% of its SM value:
\be
g_{Stt} \equiv
\sum_{j=1}^3 \frac{v v_j \mathcal{V}_{j3}
\left( V C_j^\dagger C_j V^\dagger \right)_{33}}{2 m_t^2} = 1 \pm 0.1.
\label{guitp}
\ee
%
We have also verified what constraints might
arise from the limits on $b \rightarrow s \gamma$ observations. These would arise solely
from the charged-scalars interactions, and should be similar in form to those found in
2HDM type-II, due to bottom and top quarks getting their masses from two different doublets.
Due to the mixing of the two charged scalars, we have verified that the couplings of the lightest
charged state to the fermions are suppressed compared to those one would obtain in a type-II 2HDM.
As a result, the constraints from $b \rightarrow s \gamma$ obtained for our model are much less stringent
than those found for the 2HDM type-II. Since we further consider only high masses for the charged
scalars (above 600 GeV), the constraints will be even less relevant.

We have found that,
for each and every set of parameters of the scalar potential
that has been used to produce figures~\ref{fig:v1mh2}--\ref{h1h2},
it is possible to find values for $f_1,\ldots,f_9$
which lead to observables
satisfying equations~\eqref{vusyp} and~\eqref{guitp} almost perfectly.
Indeed,
most of the observables can be fitted at the 1\,$\sigma$ level,
but at least one of the observables $m_c$,
$\left| V_{ub} \right|$,
and $J$ can only be fitted at the 2\,$\sigma$ level.
Thus,
if all the observables except $m_c$
are within their $1\, \sigma$ allowed domains,
then the minimum pull\footnote{As usual,
we define the `pull' of an observable as
the difference between its fitted value and its mean value
divided by the standard deviation.}
of $m_c$ is $1.9$;
if all the observables except $\left| V_{ub} \right|$
have pull smaller than one in modulus,
then $\left| V_{ub} \right|$ has a pull of at least $2.4$;
if all the observables but $J$ are within their $1\, \sigma$ boundaries,
then $J$ has a pull smaller than $-1.8$.
Altogether,
the best fits that we were able to achieve have a value of $\chi^2$---for
the three quark masses and the four CKM-matrix observables
in equations~(\ref{vusyp})---of 5.1.

An example of one of our best fits is provided in table~\ref{bestfit}.
\begin{table}[ht!]
\renewcommand{\arraystretch}{1.25}
\centering
\begin{tabular}{||c|c||c|c||}
\hline \hline
parameter & value & observable & value
\\ \hline
$\beta$\,(rad) & 1.5732963215865827 & $m_{h_3}$\,(GeV) & 125.0
\\ \hline
$\lambda_1$ & 17.135112092706517 & $m_{h_2}$\,(GeV) & 739.2
\\ \hline
$\lambda_2$ & 0.13092447205288404 & $m_{h_1}$\,(GeV) & 951.7
\\ \hline
$\lambda_3$ & 15.624853371379327 & $m_{A_2}$\,(GeV) & 1106
\\ \hline
$\lambda_4$ & $-11.846787927249578$ & $m_{A_1}$\,(GeV) & 1281
\\ \hline
$\lambda_5$ & 19.99999999813406 & $m_{H_2^+}$\,(GeV) & 773.0
\\ \hline
$\lambda_6$ & 16.363914697200098 & $m_{H_1^+}$\,(GeV) & 1193
\\ \hline
$\lambda_7$ & $-9.030984509839026$ & $g_{SVV}$ & 0.9925
\\ \hline
$\lambda_8$ & $-2.6314236783145977$ & $g_{Sbb}$ & 1.000
\\ \hline
$\mu_4$\,(GeV$^2$) & 2140.7424941612453 & $g_{S\tau\tau}$ & 1.000
\\ \hline
$f_1$ & 0.0024483113150037543 & $T$ & 0.01000
\\ \hline
$f_2$ & $-0.03085374618190331$ & $m_u$\,(MeV) & 1.665
\\ \hline
$f_3$ & $-0.20462462612388946$ & $m_c$\,(MeV) & 679.1
\\ \hline
$f_4$ & $0.0032554425959401188$ & $m_t$\,(GeV) & 170.9
\\ \hline
$f_5$ & 0.00697096592835829 & $\left| V_{us} \right|$ & 0.2259
\\ \hline
$f_6$ & $-0.001672825126610988$ & $\left| V_{cb} \right|$ & 0.04144
\\ \hline
$f_7$ & $-0.040296075343726166$ & $\left| V_{ub} \right|$ & 0.003694
\\ \hline
$f_8$ & 0.7561059024727611 & $J$ & 0.00002706
\\ \hline
$f_9$ & $-0.6259667363570083$ & $g_{Stt}$ & 0.9926
\\ \hline \hline
\end{tabular}
\caption{The values of the parameters and of the observables
for one of our fits.
The sign of $v_2$---see equation~\eqref{eq:vevs}---is positive for this fit.
The values of $\mu_{1,2,3}$ were computed
by using the stationarity equations for the vacuum.}
\label{bestfit}
\end{table}

\section{Conclusions}
\label{summary}

In this paper we have shown that it is possible to unify the idea
of a scotogenic neutrino mass model~\cite{ma2006}
with the enforcement of co-bimaximal lepton mixing.
The latter is obtained via softly broken lepton numbers~\cite{GL2003}
and a non-standard $CP$ transformation
which interchanges the $\mu$ and $\tau$ flavours.
Such a $CP$ transformation procures $\theta_{23} = 45^\circ$
and $\delta = \pm \pi/2$ in the lepton mixing matrix,
while $\theta_{13}$ remains undetermined;
this is in good agreement with the data.
In a scotogenic model,
the neutrino masses are generated through a one-loop diagram
involving the dark sector,
which consists of right-handed neutrinos
and a scalar gauge doublet $\eta$ which has zero VEV.
Thus,
a scotogenic model combines neutrino-mass suppression
through the seesaw mechanism and through radiative mass generation.

Our model contains three scalar doublets with nonzero VEVs.
Therefore,
we wanted to demonstrate that a scalar $h_3$
with mass $125\, \mathrm{GeV}$ can be accommodated in our model.
We have shown that this scalar can be made to have couplings
to the gauge bosons and to the heavy fermions
very close to those of the Higgs particle.
Since the non-standard $CP$ transformation interchanges,
besides the $\mu$ and $\tau$ flavours,
also two of the scalar doublets,
it is non-trivial to make all the scalars other than $h_3$ heavy.
Still,
we have found that
all of them can be made to have masses above $600\, \mathrm{GeV}$.

We have also demonstrated that the symmetries of our model
may consistently be extended to the quark sector,
correctly reproducing all the quark masses and the CKM matrix.
There are  neutral scalar-mediated flavour changing currents;
however,
it is possible to choose the model's symmetries
so that they occur only in the up-type-quark sector,
for which the experimental constraints on such currents are much looser.
The fit to the quark sector is at the 2\,$\sigma$ level,
but with many observables falling
within their 1\,$\sigma$ uncertainty intervals.

\paragraph{Acknowledgements:}
We thank Igor P.\ Ivanov for helpful discussions
concerning the scalar potential and Kristjan Kannike for valuable help
in the derivation of the conditions~(\ref{kkannike}).
D.J.\ thanks the Lithuanian
Academy of Sciences for support through the project DaFi2016.
The work of L.L.\ is supported by
the Portuguese \textit{Funda\c{c}\~ao para a Ci\^encia e a
Tecnologia} through the projects CERN/FIS-NUC/0010/2015
and UID/FIS/00777/2013;
those projects are partially funded through POCTI (FEDER),
COMPETE,
QREN,
and the European Union.

\setcounter{equation}{0}
\renewcommand{\theequation}{A\arabic{equation}}
\appendix
\section{One-loop neutrino mass corrections}

In this appendix we collect some formulae from ref.~\cite{GL2002},
adapting them to the model in this paper.
In particular,
we set $n_L = n_R = 3$ and $n_H = 4$,
where $n_L$,
$n_R$,
and $n_H$ are,
in the notation of ref.~\cite{GL2002},
the numbers of fermion families,
of right-handed neutrino singlets,
and of scalar doublets,
respectively.
The Yukawa Lagrangian of the right-handed neutrinos
in equation~(1) of ref.~\cite{GL2002} is given by
\be
\mathcal{L}_{\nu_R\, \mathrm{Yukawa}} = - \overline{\nu_R} \left(
\sum_{k=1}^4 \tilde \phi_k^\dagger\, \Delta_k \right) D_L
+ \mathrm{H.c.}
\label{Yukawa}
\ee
The notation for the physical neutral scalars
is best explained in ref.~\cite{osland}.
The neutral component of the scalar doublet $\phi_k$
($k = 1, 2, 3, 4$)
has VEV $\left\langle 0 \left| \phi_k^0 \right| 0 \right\rangle
= v_k \left/ \sqrt{2} \right.$ and is written as
\be
\label{gisfy}
\phi_k^0 = \frac{1}{\sqrt{2}}
\left( v_k + \sum_{b=1}^8 \mathcal{V}_{kb} S^0_b \right),
\ee
where the complex
matrix $\mathcal{V}$ is $4 \times 8$.
The neutral Goldstone boson is $S^0_1$
and the remaining seven $S^0_b$,
for $b = 2, \ldots, 8$,
are physical neutral scalars with masses $m_b$.
For each physical neutral scalar we define,
following ref.~\cite{GL2002},
the matrix
\be
\label{hopyi}
\Delta_b \equiv \sum_{k=1}^4 \mathcal{V}_{kb} \Delta_k.
\ee
Then,
the final result in equation~(53) of ref.~\cite{GL2002} is
\bs
\label{final1}
\ba
\delta \mnu &=& \sum_{b = 2}^8 \frac{m_b^2}{32\pi^2}\, \Delta_b^T W^\ast
\left( \frac{1}{\widetilde m} \ln{\frac{\widetilde m^2}{m_b^2}} \right)
W^\dagger \Delta_b
\label{sfoyp} \\ & &
+ \frac{3 g^2}{64 \pi^2 c_w^2}\, M_D^T W^\ast
\left( \frac{1}{\widetilde m} \ln{\frac{\widetilde m^2}{m_Z^2}} \right)
W^\dagger M_D.
\label{woypf}
\ea
\es
The sum in line~(\ref{sfoyp}) includes only the physical neutral scalars.
Line~(\ref{woypf}) includes the contributions from the loop with a $Z^0$
and from the loop with a neutral Goldstone boson.
In that line,
$M_D = \sum_{k=1}^4 \left( v_k \left/ \sqrt{2} \right. \right) \Delta_k$
is the Dirac neutrino mass matrix;
in the model in this paper that matrix vanishes,
because both the matrices $\Delta_{1,2,3}$ and the VEV $v_4$ are null.
Therefore,
for this paper only line~(\ref{sfoyp}) matters.

In equation~(\ref{final1}),
the $3 \times 3$ unitary matrix $W$
is the one that diagonalizes $M_R$
according to equation~(51) of ref.~\cite{GL2002}:
\be
\label{uidto}
W^\dagger M_R W^\ast = \widetilde m \equiv \diag \left( m_4, m_5, m_6 \right),
\ee
where $m_{4,5,6}$ are the masses of the physical heavy neutrinos.


\begin{thebibliography}{99}

\bibitem{GL2003}
W.~Grimus and L.~Lavoura,
\textit{A non-standard CP transformation
leading to maximal atmospheric neutrino mixing},
Phys.\ Lett.\ B {\bf 579} (2004) 113
[{\tt hep-ph/0305309}].

\bibitem{Forero:2014bxa}
M.~C.~Gonzalez-Garcia, M.~Maltoni and T.~Schwetz,
\textit{Updated fit to three neutrino mixing: status of leptonic CP violation},
JHEP {\bf 1411} (2014) 052
[{\tt arXiv:1409.5439 [hep-ph]}];
\\
D.~V.~Forero, M.~T\'ortola, and J.~W.~F.~Valle,
\textit{Neutrino oscillations refitted},
Phys.\ Rev.\ D {\bf 90} (2014) 093006
[{\tt arXiv:1405.7540 [hep-ph]}];
\\
F.~Capozzi, G.~L.~Fogli, E.~Lisi, A.~Marrone, D.~Montanino, and A.~Palazzo,
\textit{Status of three-neutrino oscillation parameters, circa 2013},
Phys.\ Rev.\ D {\bf 89} (2014) 093018
[{\tt arXiv:1312.2878 [hep-ph]}];
\\
J.~Bergstr\"om, M.~C.~Gonzalez-Garcia, M.~Maltoni and T.~Schwetz,
\textit{Bayesian global analysis of neutrino oscillation data},
JHEP {\bf 1509} (2015) 200
[{\tt arXiv:1507.04366 [hep-ph]}].

\bibitem{ma2015}
E.~Ma,
\textit{Neutrino mixing: $A_4$ variations},
Phys.\ Lett.\ B {\bf 752} (2016) 198
[{\tt arXiv:1510.02501 [hep-ph]}].

\bibitem{rode}
A.~S.~Joshipura and K.~M.~Patel,
\textit{Generalized $\mu$--$\tau$ symmetry and discrete subgroups of $O(3)$},
Phys.\ Lett.\  B {\bf 749} (2015) 159
[{\tt arXiv:1507.01235 [hep-ph]}];
\\
H.-J-~He, W.~Rodejohann, and X.-J.~Xu,
\textit{Origin of constrained maximal $CP$ violation in flavor symmetry},
Phys.\ Lett.\ B {\bf 751} (2015) 586
[{\tt arXiv:1507.03541 [hep-ph]}].

\bibitem{fukuura}
K.~Fukuura, T.~Miura, E.~Takasugi, and M.~Yoshimura,
\textit{Large $CP$ violation, large mixings of neutrinos,
and a democratic-type neutrino mass matrix},
Phys.\ Rev.\ D {\bf 61} (2000) 073002
[{\tt hep-ph/9909415}]; \\
T.~Miura, E.~Takasugi, and M.~Yoshimura,
\textit{Large $CP$ violation, large mixings of neutrinos,
and the $Z_3$ symmetry},
Phys.\ Rev.\ D {\bf 63} (2001) 013001
[{\tt hep-ph/0003139}]; \\
P.~F.~Harrison and W.~G.~Scott,
\textit{$\mu$--$\tau$ reflection symmetry in lepton mixing
and neutrino oscillations},
Phys.\ Lett.\ B {\bf 547} (2002) 219
[{\tt hep-ph/0210197}].

\bibitem{GL2002}
W.~Grimus and L.~Lavoura,
\textit{One-loop corrections to the seesaw mechanism
in the multi-Higgs-doublet standard model},
Phys.\ Lett.\ B {\bf 546} (2002) 86
[{\tt hep-ph/0207229}].

\bibitem{ma2006}
E.~Ma,
\textit{Verifiable radiative seesaw mechanism of neutrino mass
and dark matter},
Phys.\ Rev.\ D {\bf 73} (2006) 077301
[{\tt hep-ph/0601225}].

\bibitem{ma2016}
E.~Ma,
\textit{Soft $A_4 \to Z_3$ symmetry breaking and cobimaximal neutrino mixing},
 Phys.\ Lett.\ B {\bf 755} (2016) 348
[{\tt arXiv:1601.00138 [hep-ph]}].

\bibitem{barbieri}
R.~Barbieri, L.~J.~Hall and V.~S.~Rychkov,
\textit{Improved naturalness with a heavy Higgs: An alternative road to LHC
  physics},
Phys.\ Rev.\ D {\bf 74} (2006) 015007
[{\tt hep-ph/0603188}].

\bibitem{small_ratio}
W.~Grimus and L.~Lavoura,
\textit{Maximal atmospheric neutrino mixing
and the small ratio of muon to tau mass},
J.\ Phys.\ G {\bf 30} (2004) 73
[{\tt hep-ph/0309050}].

\bibitem{real}
L.~Lavoura,
\textit{Real CP violation in a simple extension of the standard model},
Phys.\ Rev.\ D {\bf 61} (2000) 076003
[{\tt hep-ph/9909275}];
\\
A.~Masiero and T.~Yanagida,
\textit{Real CP violation}
[{\tt hep-ph/9812225}].

\bibitem{osland}
W.~Grimus, L.~Lavoura, O.~M.~Ogreid and P.~Osland,
\textit{A precision constraint on multi-Higgs-doublet models},
J.\ Phys.\ G {\bf 35} (2008) 075001
[{\tt arXiv:0711.4022 [hep-ph]}].

\bibitem{Ivanov:2006yq}
 I.~P.~Ivanov,
\textit{Minkowski space structure of the Higgs potential in 2HDM},
Phys.\ Rev.\  D {\bf 75} (2007) 035001
[Erratum-ibid.\  D {\bf 76} (2007) 039902]
[{\tt arXiv:hep-ph/0609018}].

\bibitem{Branco:2011iw}
G.~C.~Branco, P.~M.~Ferreira, L.~Lavoura, M.~N.~Rebelo, M.~Sher and
J.~P.~Silva,
\textit{Theory and phenomenology of two-Higgs-doublet models},
Phys.\ Rept.\  {\bf 516} (2012) 1
[{\tt arXiv:1106.0034 [hep-ph]}].

\bibitem{kannike}
K.~Kannike,
\textit{Vacuum stability conditions from copositivity criteria},
Eur.\ Phys.\ J.\ C {\bf 72} (2012) 2093
[{\tt arXiv:1205.3781 [hep-ph]}].

\bibitem{Agashe:2014kda}
K.~A.~Olive {\it et al.} [Particle Data Group Collaboration],
\textit{Review of Particle Physics},
Chin.\ Phys.\ C {\bf 38} (2014) 090001.

\bibitem{LHC}
The ATLAS and CMS Collaborations,
\textit{Measurements of the Higgs boson production and decay rates
and constraints on its couplings from a combined ATLAS and CMS analysis
of the LHC pp collision data at $\sqrt{s} = 7$ and $8\, \mathrm{TeV}$},
ATLAS-CONF-2015-044.

\bibitem{Xing:2011aa}
Z.~z.~Xing, H.~Zhang and S.~Zhou,
\textit{Impacts of the Higgs mass on vacuum stability,
running fermion masses, and two-body Higgs decays},
Phys.\ Rev.\ D {\bf 86} (2012) 013013
[{\tt arXiv:1112.3112 [hep-ph]}].

\end{thebibliography}
\end{document}